\documentclass[showpacs,prd,reprint,preprintnumbers,amsmath,amssymb,superscriptaddress,nofootinbib]{revtex4-1}

\usepackage{graphicx,subfigure,multirow}
 \usepackage{longtable}
\usepackage{dcolumn}
\usepackage{bm}
\usepackage[all]{xy}
\usepackage{color}
\usepackage{slashed}
\usepackage[usenames,dvipsnames]{xcolor}
\usepackage[unicode=true,pdfusetitle,
 bookmarks=true,bookmarksnumbered=false,bookmarksopen=false,citecolor=Turquoise,
 breaklinks=false,pdfborder={0 0 1},backref=false,colorlinks=true]
 {hyperref}

\usepackage{ulem}
\newcommand{\TeV}{\,\textrm{TeV}}

\newcommand{\GeV}{\,\textrm{GeV}}
\makeatletter

\newcommand{\fmslash}[2][0mu]{%
  \mathchoice
    {\fmsl@sh\displaystyle{#1}{#2}}%
    {\fmsl@sh\textstyle{#1}{#2}}%
    {\fmsl@sh\scriptstyle{#1}{#2}}%
    {\fmsl@sh\scriptscriptstyle{#1}{#2}}}
\newcommand{\fmsl@sh}[3]{%
  \m@th\ooalign{$\hfil#1\mkern#2/\hfil$\crcr$#1#3$}}
\makeatother

\newcommand{\lsim}{{\;\raise0.3ex\hbox{$<$\kern-0.75em\raise-1.1ex\hbox{$\sim$}}\;}}
\newcommand{\gsim}{{\;\raise0.3ex\hbox{$>$\kern-0.75em\raise-1.1ex\hbox{$\sim$}}\;}}

\newcommand{\beq}{\begin{equation}}
\newcommand{\eeq}{\end{equation}}
\newcommand{\bea}{\begin{eqnarray}}
\newcommand{\eea}{\end{eqnarray}}
\mathchardef\minus="002D

\usepackage{color}

\begin{document}
\title{Examining the origin of dark matter mass at colliders}
\author{Minho Kim}
\email{kmhmon@postech.ac.kr}
\affiliation{Department of Physics, POSTECH, Pohang 37673, Korea}
\affiliation{Institute of Convergence Fundamental Studies and School of Liberal Arts, \\
Seoul National University of Science and Technology, Seoul 01811, Korea.}
\author{Hye-Sung Lee}
\email{hyesung.lee@kaist.ac.kr}
\affiliation{Center for Theoretical Physics of the Universe, Institute for Basic Science (IBS), Daejeon, 34051, Korea}
\affiliation{Department of Physics, KAIST, Daejeon 34141, Korea}
\author{Myeonghun Park}
\email{parc.seoultech@seoultech.ac.kr}
\affiliation{Institute of Convergence Fundamental Studies and School of Liberal Arts, \\
Seoul National University of Science and Technology, Seoul 01811, Korea.}
\affiliation{Center for Theoretical Physics of the Universe, Institute for Basic Science (IBS), Daejeon, 34051, Korea}
\author{Mengchao Zhang}
\email{mczhang@ibs.re.kr}
\affiliation{Center for Theoretical Physics of the Universe, Institute for Basic Science (IBS), Daejeon, 34051, Korea}
\date{Aug 28, 2018}

\preprint{
CTPU-16-45
}

\begin{abstract}
As conventional dark matter scenarios have been probed extensively so far, the physics of a light dark matter charged under a new gauge group (dark gauge group) becomes one of new research avenues in many theoretical and experimental studies.
We examine properties of a dark photon showering, the radiation process of light gauge bosons from energetic dark matter particles produced at the Large Hadron Collider (LHC).
This showering process provides different signatures at the LHC depending on the property of dark matter under the dark gauge group.
We show that the LHC experiment can identify the chirality of a dark matter, which leads to understanding the mass origin of particles in the dark sector.
\end{abstract}


\maketitle

\section{Introduction}
The confirmation of dark matter (DM) existence will be the one of major milestones toward the physics beyond the standard model (SM) of the particle physics.
Among various scenarios about dark matter, a weakly-interacting massive particle (WIMP)\,\cite{Lee:1977ua}, often in the supersymmetry framework \cite{Jungman:1995df}, has been extensively tested by various dark matter direct detection (DD) experiments\,\cite{currentDD,futureDD} together with collider experiments including the LHC\,\cite{LHCDark}.
In near future, sensitivity of DD experiments will reach the point of detecting irreducible backgrounds from neutrino-nucleus coherent scattering\,\cite{Billard:2013qya}.

In contrast to WIMP dark matter, searching for a light dark matter of sub-GeV mass scale with conventional DD experiments is very challenging 
due to low nuclear recoil energy $E_{\textrm{NR}} < \mathcal{O}(0.1)\,\textrm{keV}$ over experimental resolutions and noises\,\cite{Light_DD}.
There have been growing interests in the sub-GeV dark matter recently, and new experiments have been initiated and proposed\,\cite{Battaglieri:2016ggd,Alexander:2016aln,deNiverville:2016rqh}.
They include direct searches of a relic dark matter particle\,\cite{Essig:2012yx} as well as beam experiments that produce dark matter particles and detect their signals by using low-energy beam facilities\,\cite{Dharmapalan:2012xp,Banerjee:2016tad}.
Contrast to DD experiments, the LHC has provided results of detecting a sub-GeV dark matter particle by utilizing initial state radiation (ISR) jet to tag events with dark matter particles as we have accumulated precise understanding in Quantum Chromodynamics (QCD) to suppress SM backgrounds\,\cite{LHC_LD}.

Together with conventional dark matter experiments which are sensitive mostly to interactions between dark matter and SM particles, 
it would be interesting to consider phenomenological effects of dark matter scenarios if there exists an interaction among dark matter itself\,\cite{Gradwohl:1992ue,Spergel:1999mh}. 
One of natural methods to implement an interaction among dark matter particles is to introduce a dark gauge symmetry on dark matter\,\cite{Hooper:2008im, Feng:2008ya,Ackerman:mha, Feng:2009mn, 
Hochberg:2014kqa,Kopp:2016yji}.
Especially a light dark gauge boson has been spotlighted in the intensity frontier research \cite{EXdarkPhoton}.
The combination of a light dark matter particle and a light gauge boson fits well, as a light gauge boson can provide a suitable annihilation channel for the dark matter particles demanded by the DM relic density constraint\,\cite{Boehm:2003hm,Fayet:2004bw}.

In this letter, we investigate generic collider signatures of a light dark matter particle, which is charged under a dark gauge group. The collider phenomenology of non-abelian dark gauge group has been studied for a composite dark matter particle\,\cite{darkSUN,Cohen:2015toa,Englert:2016knz}. 
Here we focus on an abelian dark gauge group, motivated by current efforts in dark photon searches\,\cite{EXdarkPhoton}.
Dark photon, as a dark gauge boson, can be produced through decay processes\,\cite{Kong:2014jwa}, or final state radiation\,\cite{decay_DP, radia_DP}. 
Dark photon shower process can be triggered, once a ``dark charged" dark matter particle is produced with a sufficient energy\,\cite{Buschmann:2015awa}.
Our emphasis is to explain the difference in collider signatures, depending on the chirality of dark matter under a dark gauge group\footnote{A systematic analysis of the dark mass origin and its impact on Cosmology and Astronomy is given in \cite{Bell:2016uhg}.}.
A chiral interaction is induced by a dark higgs boson if it provides a mass to a dark photon through dark symmetry breaking and if dark matter becomes massive via yukawa interaction.
Thus we point out that recognizing patterns in dark photon showering can be a good probe to examine the mass origin of particles in a dark sector.

\section{Dark Sector}
Here we describe a minimal dark sector with a dark matter under a dark gauge group $U(1)_d$. If a dark sector contains a dark matter as a fundamental particle, the corresponding Lagrangian will be following;
\beq
\mathcal{L}_\text{vector+scalar} \ni -\frac{1}{4} F'_{\mu\nu} F'^{\mu\nu} + \frac{\varepsilon}{2} F_{\mu\nu} F'^{\mu\nu} + | D_\mu \Phi |^2 , ~
\label{mixingL}
\eeq
\begin{eqnarray}
\mathcal{L}_\text{matter} &=& \bar\chi_L i \gamma^\mu  D_\mu \chi_L + \bar\chi_R i \gamma^\mu D_\mu \chi_R + \bar\psi_L i \gamma^\mu  D_\mu \psi_L \nonumber \\ 
&+& \bar\psi_R i \gamma^\mu D_\mu \psi_R - y_\chi \bar\chi_L \Phi^* \chi_R - y_\chi \bar\chi_R \Phi \chi_L \nonumber \\
&-& y_\psi \bar\psi_L \Phi \psi_R - y_\psi \bar\psi_R \Phi^* \psi_L , 
\label{eq:yukawa}
\end{eqnarray}
with $D_\mu \equiv \partial_\mu + i g' Q' A'_\mu$ where $A'_{\mu}$ is the quantum field of a dark photon $\gamma_d$, $g'$ is the gauge coupling of the dark gauge symmetry, $Q'$ is the dark $U(1)_d$ charge.
$F_{\mu\nu}$ and $F'_{\mu\nu}$ are the field strength of the SM photon and dark photon respectively, and $\varepsilon$ is the kinetic mixing parameter \cite{kineticmixing}. The SM particles are not charged under a dark gauge group. Here as we focus on a light dark photon where the mass of a dark photon is negligible compared to the mass of $Z$ boson, the effect of SM electroweak symmetry breaking on a kinetic mixing becomes irrelevant to interactions between a dark photon and particles in the SM\,\cite{Chun:2010ve,Lee:2016ief}. 
$\Phi$ is a dark higgs which may break $U(1)_d$ depending on its charge under the dark gauge group.
We introduce two pairs of chiral fermions $\chi$ as dark matter and $\psi$ as a heavier particle in a dark sector for the anomaly cancellation, which is model-dependent part. The yukawa terms in eq.\,\eqref{eq:yukawa} dictate the relations of the dark $U(1)_d$ charges;
\bea
Q'_{\chi_L} - Q'_{\chi_R} + Q'_\Phi &=& 0\, , \label{eq:coupling1}\\
-Q'_{\psi_L} + Q'_{\psi_R} + Q'_\Phi &=& 0\,.
\eea
For the anomaly cancellation, we take $Q'_{\chi_L/ \chi_R} = -Q'_{\psi_L / \psi_R}$, which allows a mixing between the $\chi$ and $\psi$.
In our analysis, however, we will focus on the phenomenology of a dark matter $\chi$.
As pointed out in \cite{Bell:2016uhg}, non-zero charge $Q'_\Phi$ of a dark higgs induces the chiral nature of dark matter. When we rewrite the interaction between a dark matter and a dark photon, 
\beq
\mathcal{L}_\textrm{matter} \ni  - g' Q'_V A'_\mu \bar \chi \gamma^\mu \chi - g' Q'_A  A'_\mu \bar \chi \gamma^\mu \gamma_5 \chi \, ,
\label{eq:darkint}
\eeq
the axial $Q'_A$ and vector coupling $Q'_V$ are written as;
\bea
Q'_A && = \frac{1}{2} \left(Q'_{\chi_R} -Q'_{\chi_L}\right) = \frac{Q'_\Phi}{2}\, \\
Q'_V && = \frac{1}{2} \left(Q'_{\chi_R} +Q'_{\chi_L}\right) = \frac{Q'_\Phi}{2}+Q'_{\chi_L}\, .
\label{coupling3}
\eea
The chirality of a dark matter particle ($Q'_A \ne 0$) results in a significant difference in collider signatures at the LHC as we will show later.\footnote{This can be inferred from, for example, the distinct phenomenology depending on the chirality of SM fermions on dark gauge group\,\cite{Davoudiasl:2012ag,Davoudiasl:2012qa} when corresponding dark gauge boson is very light.} Checking the chirality of a dark matter is directly related to understanding the mass generating mechanism for a dark matter. 

There are several ways to address a small dark gauge boson mass even when a gauge coupling constant is not small \cite{ArkaniHamed:2008qp,Cheung:2009qd,Lee:2016ejx}. In this paper we take a small vacuum expectation value $v_S$ of the dark higgs $\Phi$ given by $\Phi = \frac{1}{\sqrt{2}} \left( v_S + S + i \phi_S \right)$.
The masses of the dark gauge boson and dark matter are given by the vacuum expectation value of the dark higgs boson as $m_{\gamma_d} = g' Q'_{\Phi} v_S$ and $m_{\chi} = y_{\chi} v_S / \sqrt{2}$.

\subsection{Dependence of dark photon showering on a mass mechanism in a dark sector}
As an accelerated charged particle radiates corresponding gauge particles, energetic dark matter particles which are produced at a high energy collider will radiate off dark gauge bosons. 
The radiation pattern of a dark photon $\gamma_d$ from an energetic dark matter $\chi$, called the showering process, depends on the mechanism of the mass generation for a dark matter 
since dark matter couples differently with a dark photon as in eq.\,\eqref{eq:darkint}. This showering process is characterized by a splitting function $P_{\chi \to \chi \gamma_d}$ which describes an emission process. 
In a collinear region, the differential probability of the splitting process $\chi \to \chi \gamma_d$ is;
\begin{equation}
\frac{\alpha^{\prime}}{2\pi}dx\frac{dt}{t}P_{\chi \to \chi \gamma_d}(x,t).
\label{kernel}
\end{equation}
Here, $\alpha^{\prime}=g^{\prime2}/4\pi$, $t$ is the virtuality of incoming $\chi$, and $x$ is the energy fraction taken by outgoing $\chi$. 
A detailed analysis of  dark photon showers from vector-like dark fermion model has been studied in \cite{Buschmann:2015awa}. 
As we focus on the phenomenology of an energetic dark matter production at colliders, we ignore terms suppressed by $m^2_{\chi}/t$ or $m^2_{\gamma_d}/t$. In this limit, the splitting kernel for vector-like dark matter is given\,\cite{Catani:2002hc};
\begin{equation}
P_{\chi \to \chi \gamma_d}(x,t) \simeq  Q'^2_V \, \frac{1+x^2}{1-x}\, .
\label{kernel1}
\end{equation}
This shower pattern is similar to the familiar QED shower which only includes contributions from transverse polarizations of a photon. 

The longitudinal polarization vector of a dark photon will grow as $E_{\gamma_d}/m_{\gamma_d}$, with the energy of a dark photon $E_{\gamma_d}$.
This artificial enhancement can be tamed by the Goldstone boson equivalence theorem (GBET)\,\cite{Goldstone}.  
While the leading contribution of a longitudinal polarization in high energy limit is expressed by GBET which is proportional to $m_{\chi}/m_{\gamma_d}$, the remaining part is suppressed by $m_{\gamma_d}/E_{\gamma_d}$\,\cite{Gaugechoose1,Gaugechoose2,Chen:2016wkt}. Thus this sub-leading part can be neglected in our study as we are interested in the phase space region of $E_{\gamma_d} \gg m_{\gamma_d}$. 
With GBET, we obtain a splitting kernel for the chiral fermions as following;
\begin{equation}
P_{\chi \to \chi \gamma_d}(x,t) \simeq \left(Q'^2_V + Q'^2_A\right) \frac{1+x^2}{1-x} + 2Q'^2_A\,\frac{m^2_{\chi}}{m^2_{\gamma_d}}.
\label{kernel2}
\end{equation}
The first term is from the transverse modes of a dark photon. The second term is from the longitudinal mode of a dark photon, which would be significant when a dark photon is very light compared to dark matter.
Unlike the chiral dark matter, a vector-like dark matter does not have an interaction with a Goldstone boson. Thus only transverse polarization of a dark photon is involved in a showering process as in eq.\,\eqref{kernel1}.
In the next section, we show the corresponding collider phenomenology by examining dark photon showering pattern.

\section{Phenomenology of the dark showering at the LHC}

\begin{figure}[t!]
\begin{center}
\includegraphics[width=0.3\textwidth]{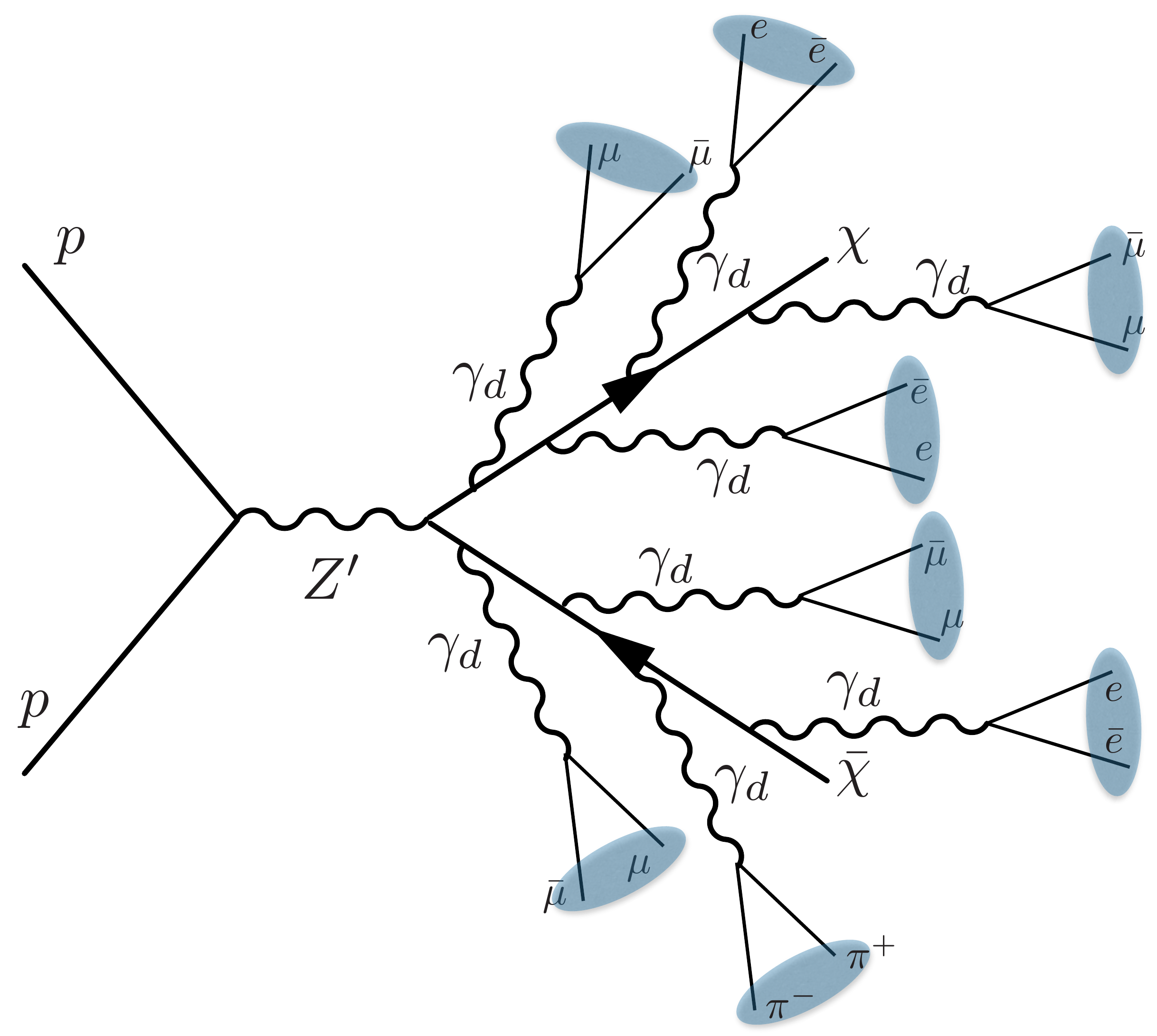}
\caption{A High energy collider produces dark matter particles, which get enough energy recoils from a decay of a heavy resonance to shower dark photons.}
\label{diagram}
\end{center}
\end{figure}
The parameters that are directly related to a dark photon showering are $\left(\alpha', m_{\chi}, m_{\gamma_d}\right)$ and $\left(Q'_A,\,Q'_V\right)$.
In a case of chiral dark matter induced by non-zero $Q'_\Phi (=1)$,  we probe the case of $Q'_{\chi_R} =0$ which maximizes effects of chirality with $Q'_{\chi_L} = -Q'_\Phi = -1$ as in eq.\,(\ref{eq:coupling1}). In terms of axial and vector coupling, we will have $\left(Q'_A,\,Q'_V\right) =\left(\frac{1}{2},\,-\frac{1}{2}\right)$ in this chiral case. For a vector-like dark mater scenario,  we consider $\left(Q'_A,\,Q'_V\right) =\left(0,\,1\right)$ which comes from $\left(Q'_{\chi_L},\,Q'_{\chi_R}\right) =\left(1, 1 \right)$ as in eq.\,\eqref{coupling3}. In this case, $Q'_\Phi$ becomes 0, which in turns decouples the origins of dark matter mass and dark photon mass.

The longitudinal component of a splitting kernel in eq.\,\eqref{kernel2} indicates that a large mass hierarchy between $m_{\chi}$ and $m_{\gamma_d}$ will induce significant difference in the shower process from the chiral dark matter compared to the case of a vector-like dark matter.
But in order to keep the dark yukawa coupling within a perturbative limit as $(y_\chi/\sqrt{2})^2 \lesssim 4\pi$, the mass spectrum of $m_{\chi}$ and $m_{\gamma_d}$ follow a limit;
\beq
\alpha^{\prime}\frac{m^2_{\chi}}{m^2_{\gamma_d}} \lesssim 1\, .
\label{limit}
\eeq
\begin{table}[t!]
\begin{tabular}{|c|c|c|c|}
\hline
Benchmark Points (BP) & ~~~A~~~ & ~~~B~~~ & ~~~C~~~ \\\hline
$\alpha^{\prime}$ & 0.3 & 0.15 & 0.075 \\\hline
$m_{\chi} \,(\GeV)$& 0.7 & 1.0 & 1.4 \\\hline
$m_{\gamma_d} \, (\GeV)$ &  \multicolumn{3}{c|}{0.4}\\\hline
\end{tabular}
\caption{Benchmark points we have chosen.
They obey the perturbative limit of  $\alpha^{\prime}\frac{m^2_{\chi}}{m^2_{\gamma_d}} \lesssim1$.}
\label{tab:BP}
\end{table}

To avoid constraints from current dark matter and dark photon search experiments\,\cite{currentDD, futureDD,Light_DD, Essig:2012yx,Dharmapalan:2012xp,  Banerjee:2016tad, Battaglieri:2016ggd, deNiverville:2016rqh} we take quite light benchmark points as in Table\,\ref{tab:BP}.
In our benchmark points of $m_{\gamma_d} = 0.4\GeV$, a viable kinematic mixing parameter $\epsilon$ can provide prompt decays of dark photons to the SM particles.
More specifically, for our bench mark points, $\epsilon^2$ smaller than $10^{-7}$ is still allowed by current constraints\,\cite{Alexander:2016aln}. 
If $\epsilon$ is too small, SM particles from a dark photon decays would leave displaced vertices for us to enhance the search power at the LHC \cite{Aad:2014yea}.
With $\epsilon^2 > 10^{-10}$, the impact parameter of particles from dark photon is smaller than 1\,mm, as particles from dark photon decays can be treated as prompt\,\cite{Aad:2014fxa}.
In this case, a dark photon mostly decays into a pair of light leptons where these non-conventional signatures are easy to tag over the QCD backgrounds. 
The corresponding branching ratios are BR$(\gamma_d\to \mu^+\,\mu^-) \simeq0.45$, BR$(\gamma_d\to e^+\,e^-) \simeq 0.45$, and BR$(\gamma_d\to \pi^+\,\pi^-) \simeq0.10$.
From now on, we fix the mass of $\gamma_d$ to $0.4$GeV.

Particles from dark photon showering processes would be tagged at collider detectors when they can leave certain level of energy deposits. 
Since a dark photon showering process as a final state radiation of dark matter is insensitive to a production process, we consider a TeV-scale mediator $Z'$ in producing dark matter particles at a collider.  As null results of the LHC push the possible mass range of a mediator to be heavy, our set up in the framework of ``hidden valley"\,\cite{Strassler:2006im} is empirically supported.\footnote{Considering a very heavy $Z'$ can be introduced as a mechanism to make $\gamma_d$ very light through  mass a matrix diagonalization\,\cite{Lee:2016ejx}.}
For a phenomenological study in a hadron collider, we take minimal interactions between the Standard Model sector and a dark sector when the mass of $Z'$ is within the coverage of the high luminosity (HL) of the LHC;
\beq
\mathcal{L} \ni - g_q Z'_\mu \bar{q} \gamma^{\mu} q +  g_\chi Z'_\mu \bar{\chi} \gamma_{\mu} \chi\, .
\label{zprime}
\eeq
here $q$ denotes a SM quark. 
If the energy of a hadron collider is not enough to produce a on-shell mediator, dark matter productions will be described by an effective operator with a mediator being integrated out\,\cite{Goodman:2010ku}.
\beq
\mathcal{L} \ni \frac{1}{\Lambda^2} (\bar{q} \gamma^{\mu} q) (\bar{\chi} \gamma_{\mu} \chi)\, .
\label{effL}
\eeq
In this case, a phase space of events with boosted dark matter is different from our current study as an initial state radiation jet will be the source of producing boosted dark matters. Here we focus on a situation where HL-LHC can reach the mass range of a mediator with $M_{Z'} = 1.5\TeV$. 
Model parameters of coupling constants are chosen to be compatible with current LHC searches of dijet and prompt lepton-jet searches as we show later.
We perform Monte Carlo studies with FeynRules\,2.0\,\cite{Alloul:2013bka} to implement dark matter models, MadGraph\_aMC@NLO\,\cite{Alwall:2014hca} and Pythia\,8\,\cite{pythia1}. 
In simulating dark photon showering processes, we modify a Hidden Valley model\,\cite{Strassler:2006im, Han:2007ae} implemented in Pythia\,8\,\cite{pythiaHV, pythia2} to add longitudinal term in eq.\,\eqref{kernel2}.

\begin{figure}[t!]
\centering
\includegraphics[width=0.232\textwidth]{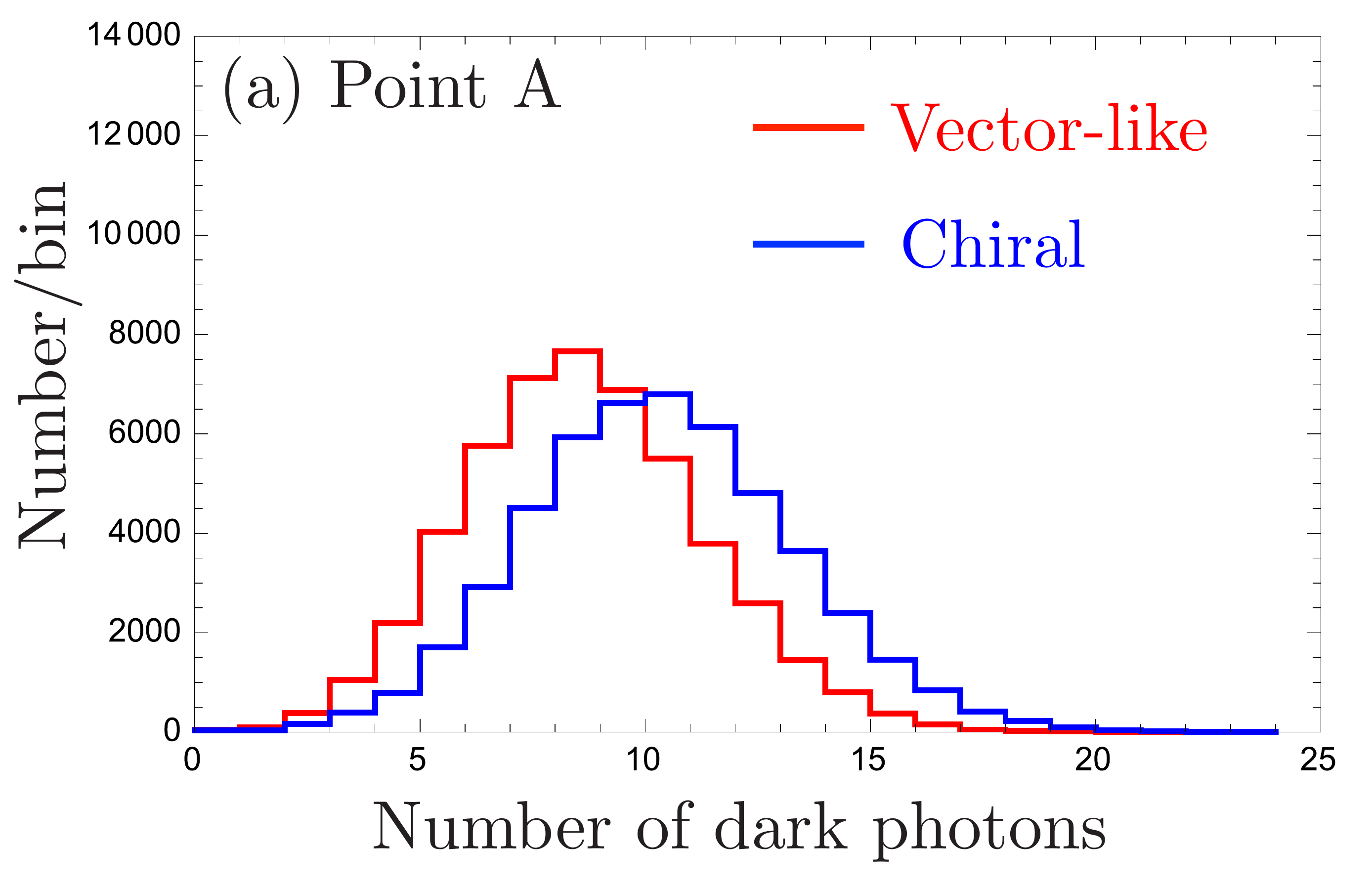} \,
\includegraphics[width=0.232\textwidth]{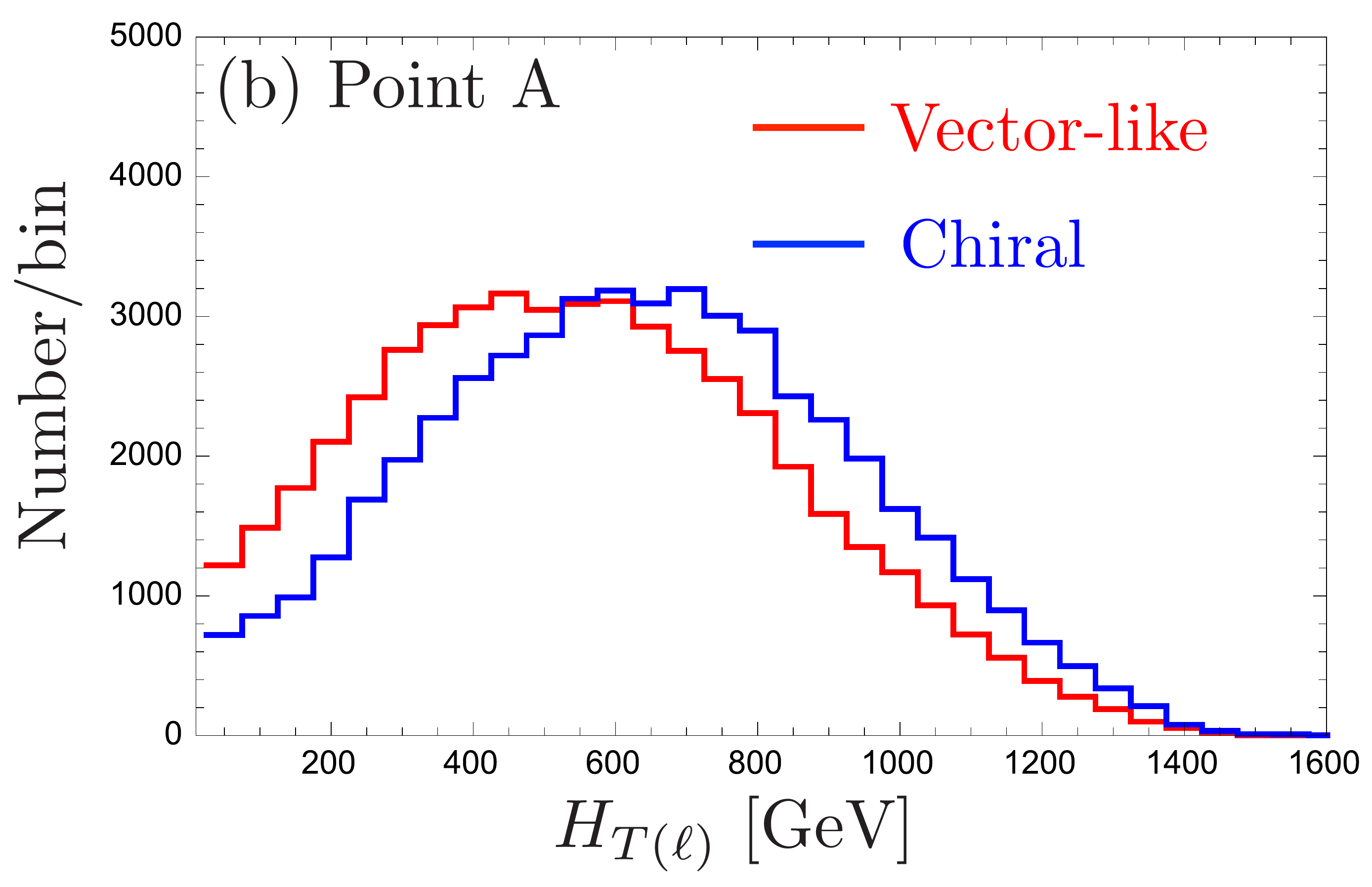} \\
\includegraphics[width=0.232\textwidth]{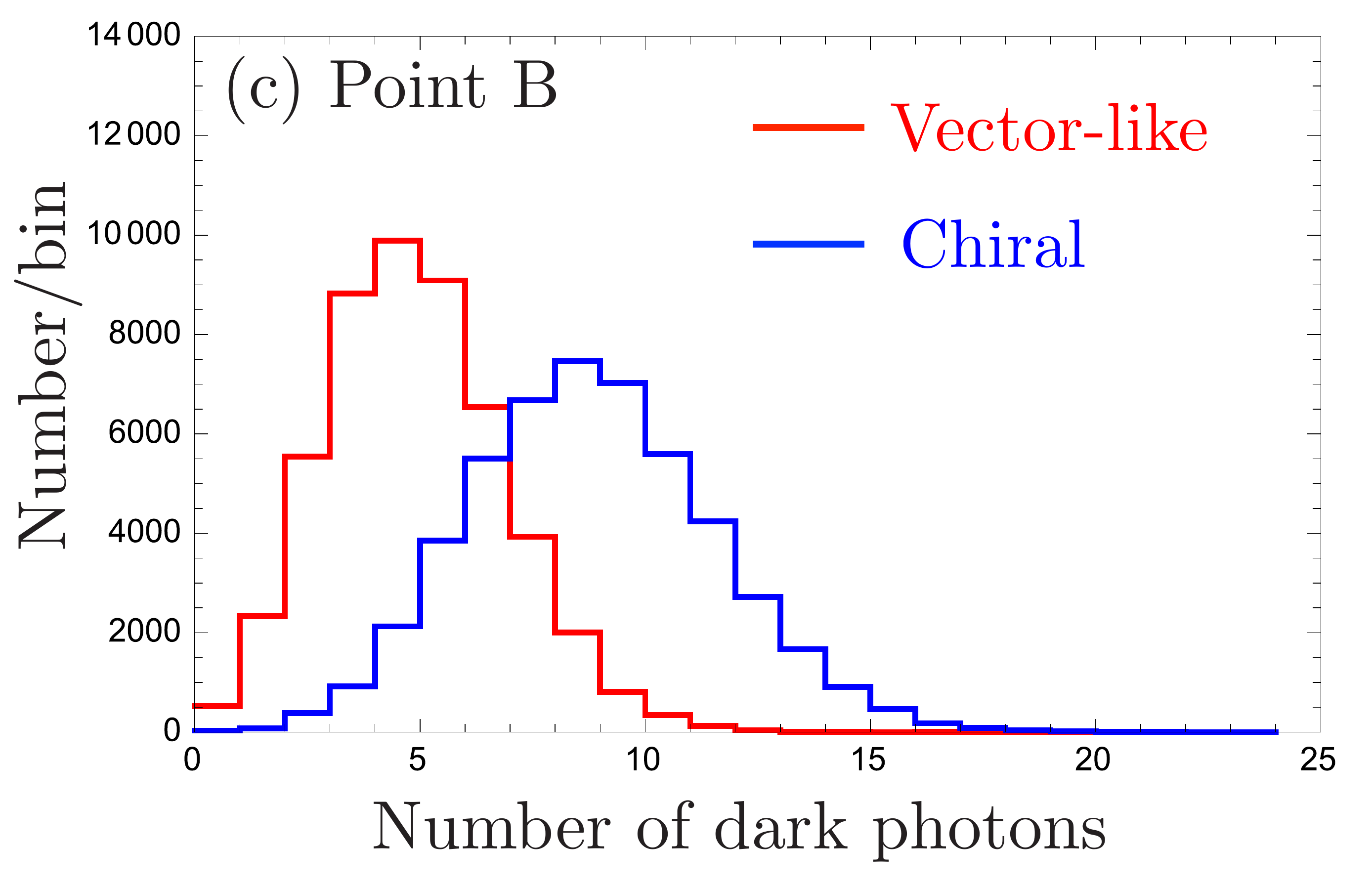} \,
\includegraphics[width=0.232\textwidth]{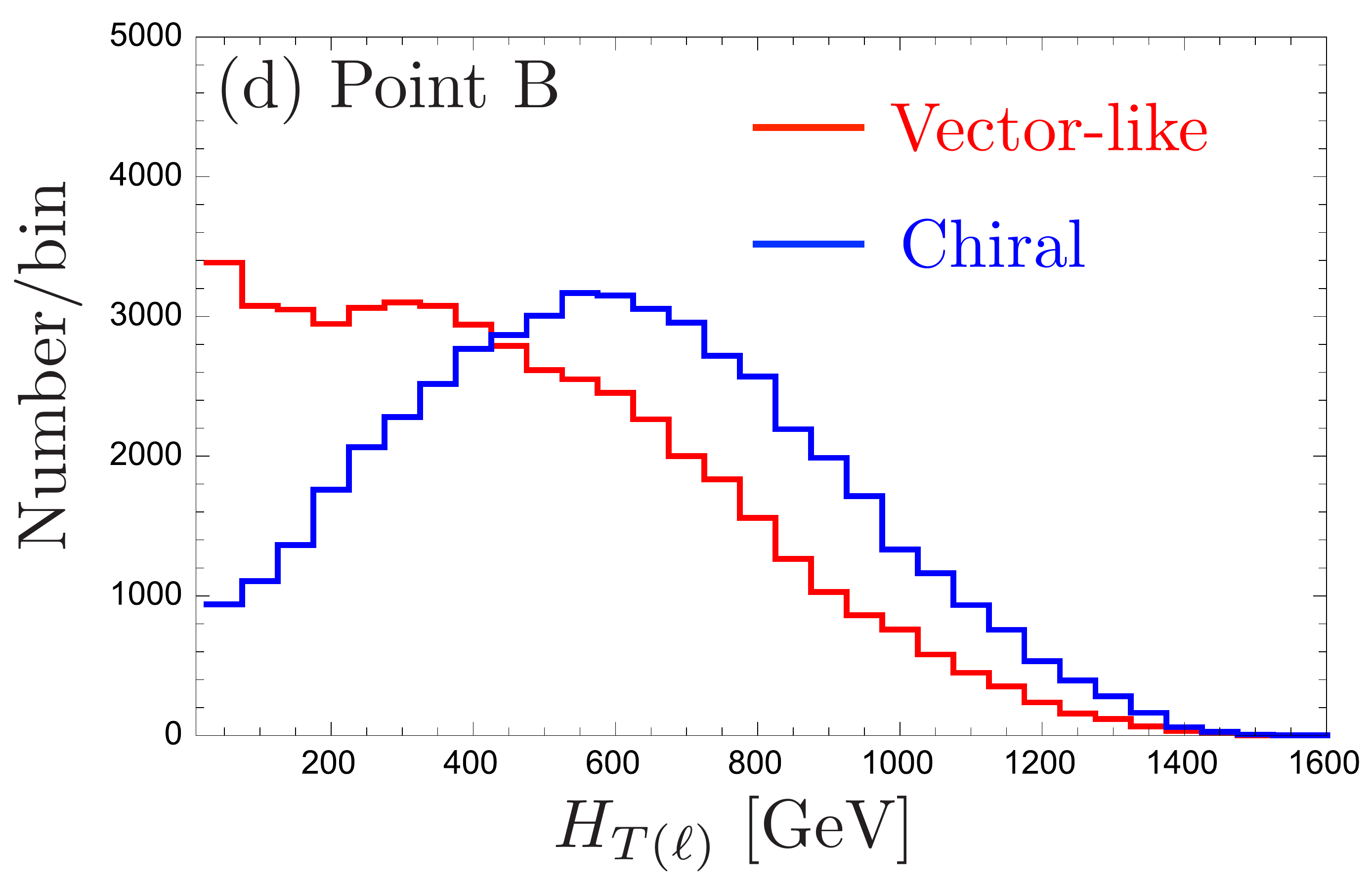} \\
\includegraphics[width=0.232\textwidth]{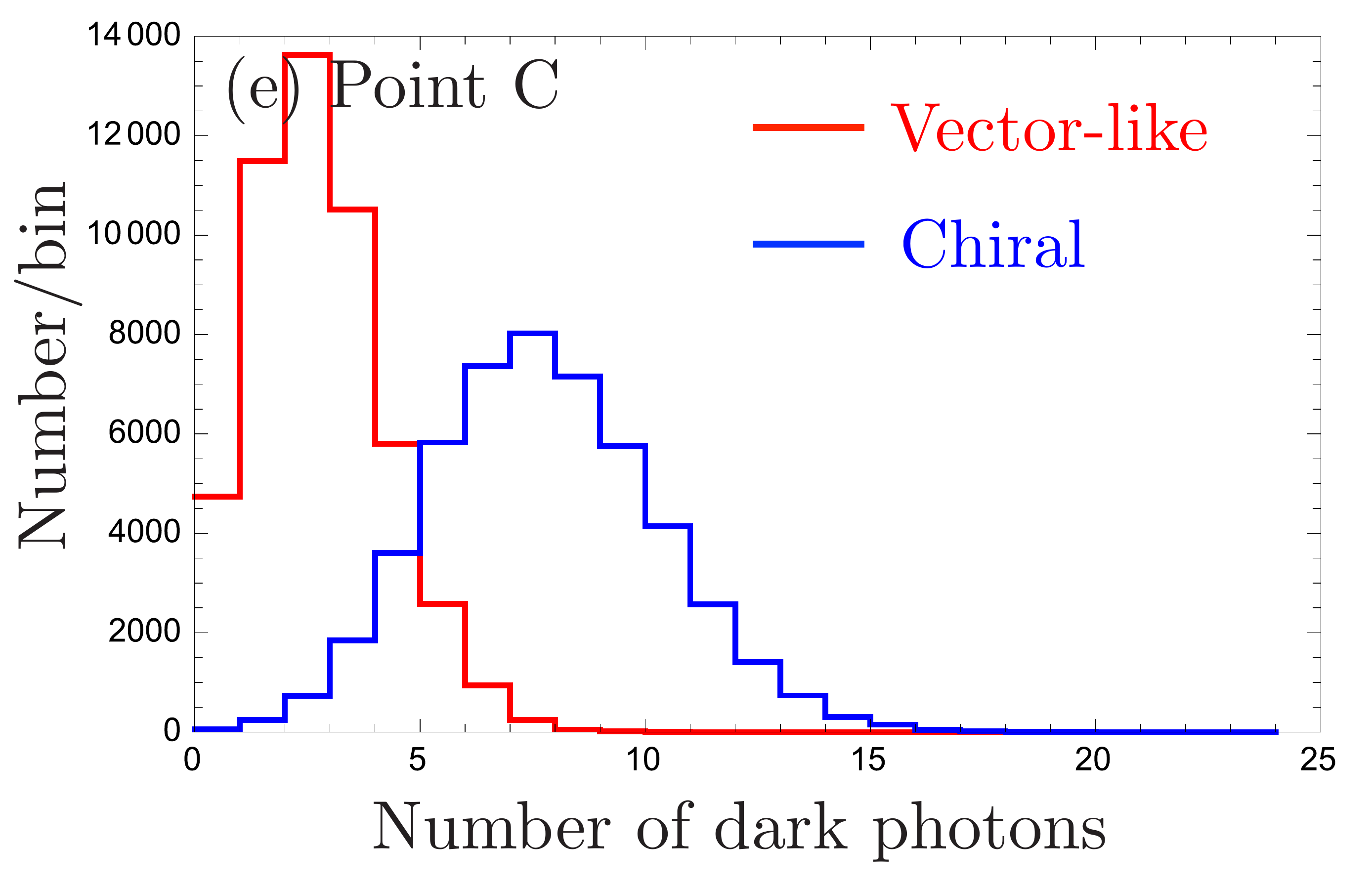}\,
\includegraphics[width=0.232\textwidth]{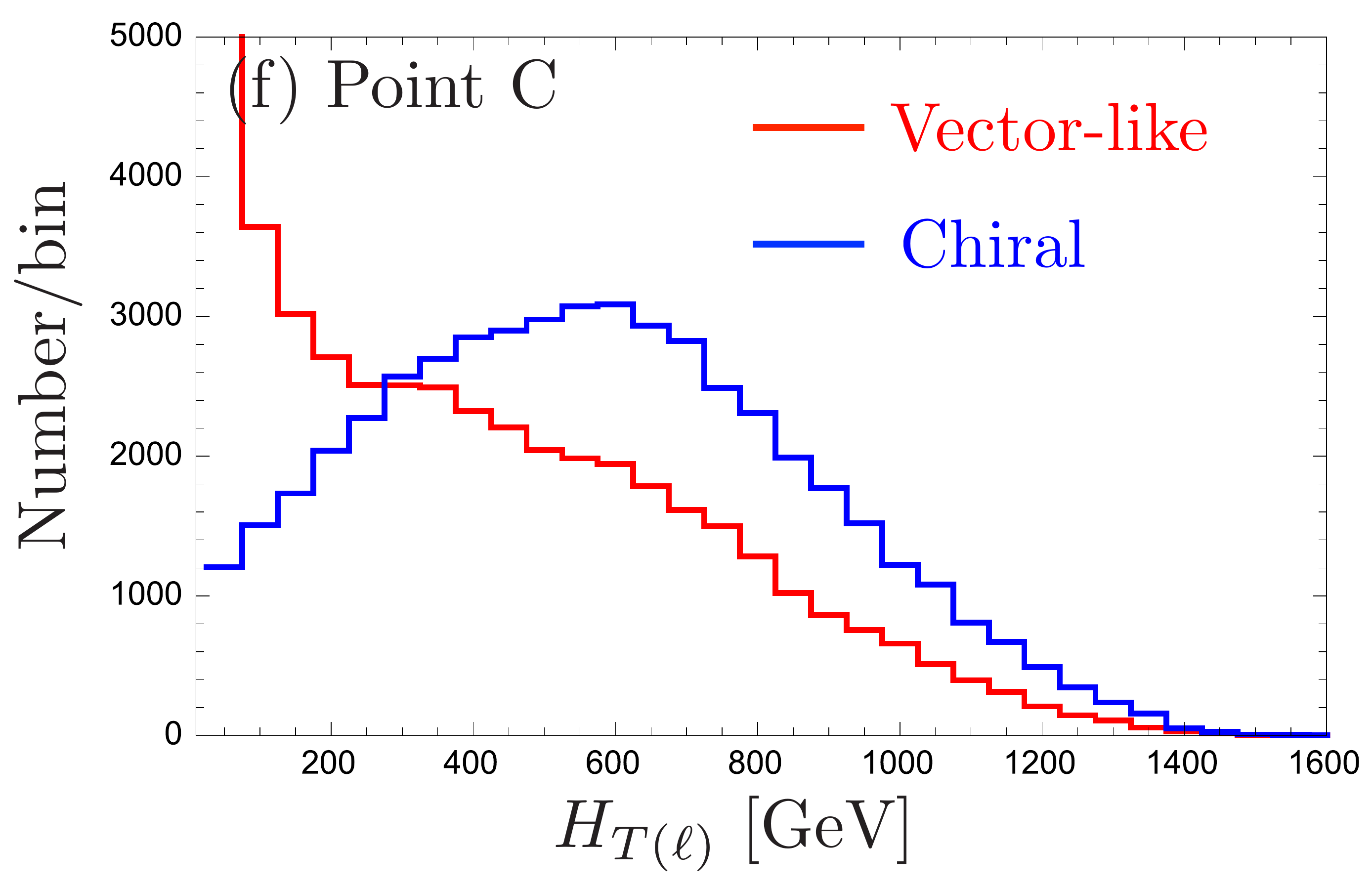}
\caption{\label{DPnumber} To show different dark photon shower pattern, we demonstrate the number of showered dark photons and corresponding $H_{T(\ell)}$ distributions with parton level Monte Carlo simulations. The left column is for the comparison of the dark photon number distribution of our benchmark points between a vector-like dark matter and chiral dark matter. Red line and blue line correspond to vector-like model and chiral model respectively. The right column is for the comparison of the ${H_T}_{(\ell)}$ distribution of our benchmark points with the same color code as cases in the left column.}
\end{figure}

To examine a difference in signatures at the LHC from distinct showering patterns, we check how many changes occur in the number of produced dark photons at the LHC depending on the chirality of dark matter. 
In the left column of Fig.\,\ref{DPnumber}, we plot histograms of the number of showered dark photons per event.
As we observe, with increasing the mass of dark matter, the number of showered dark photons is reduced in a vector-like dark matter case as a dark gauge coupling $\alpha^{\prime}$ is decreasing as in our benchmark points in Tab.\,\ref{tab:BP}.
But in a chiral dark matter scenario, the number of the dark photon is almost unchanged due to the enhancement from the GBET with a large yukawa coupling in the second term of eq.\,\eqref{kernel2}.

Once, a dark photon is produced from a dark showering process, it decays to SM particles through a kinetic mixing $\epsilon$ in eq.\,(\ref{mixingL}).
To quantify those signatures, we use the scalar sum of the muon and electron transverse momentum $p_T$;
\begin{equation}
{H_T}_{(\ell)} = \sum_{i = \mu^\pm, e^\pm } |{p_T}_i\,|.
\label{HTsum}
\end{equation}
In the right column of Fig.\,\ref{DPnumber} we observe that a dark photon shower in the vector-like dark matter case becomes weak and most of leptons become soft as a coupling constant $\alpha^{\prime}$ gets smaller. 
In a chiral dark matter case, showering processes with a longitudinal mode of a dark photon, which is independent on $\alpha^{\prime}$, becomes dominant through an enhancement from a yukawa interaction as dark matter becomes massive. This can be shown in the following limit;
\beq
\lim_{\alpha' \ll 1}\alpha' P_{\chi \to \chi \gamma_d}\simeq \alpha' \cdot 2Q_A'^2\frac{m_\chi^2}{m_{\gamma_d}^2} \propto \frac{m_{\gamma_d}^2}{v_S^2}\cdot \frac{m_\chi^2}{m_{\gamma_d}^2} \sim y_\chi^2 \, .
\eeq 
Due to GBET, the energy spectrum of leptons from a longitudinal mode is larger compared to the case of leptons from a transverse mode of dark photons.

To consider various factors including isolation of reconstructed objects and smearing effects on energy deposits, we use Delphes 3\,\cite{deFavereau:2013fsa} as a fast detector simulation with the ATLAS parameter setting.
We adopt the concept of lepton-jet (LJ)\,\cite{Aad:2014yea,leptonJet} to cluster collimated muons from a light dark photon.
As a light dark photon decays mostly the pair of leptons, we only consider a muon-jet ($\textrm{LJ}_\mu$) as a candidate to suppress QCD backgrounds as we can identify individual muons in $\textrm{LJ}_\mu$ by utilizing various sub-detectors to perform $\chi^2$ fitting for muon tracks. 
Considering electron-jet ($\textrm{LJ}_e$) for a signal object would non-trivial as bremsstrahlung processes of electrons force us to merge several crystals in electronic calorimeters, making it difficult to isolate each electrons in collimated situations and requiring additional properties to reduce QCD backgrounds\,\cite{Khachatryan:2015hwa,Aad:2014yea}.
Thus we require at least two muon-jets ($\textrm{LJ}_\mu$) to tag signals over backgrounds.
After we require an isolation criteria for a muon-jet, we choose the mass of muon-jet ($\textrm{LJ}_\mu$) within the range between $0.3\GeV$ and $0.5\GeV$ to consider imperfect resolution of detectors\,\cite{Aad:2016jkr}. 

\subsection{Constraints from the LHC searches}
\begin{figure*}[th!]
\centering
\includegraphics[width=0.4\textwidth]{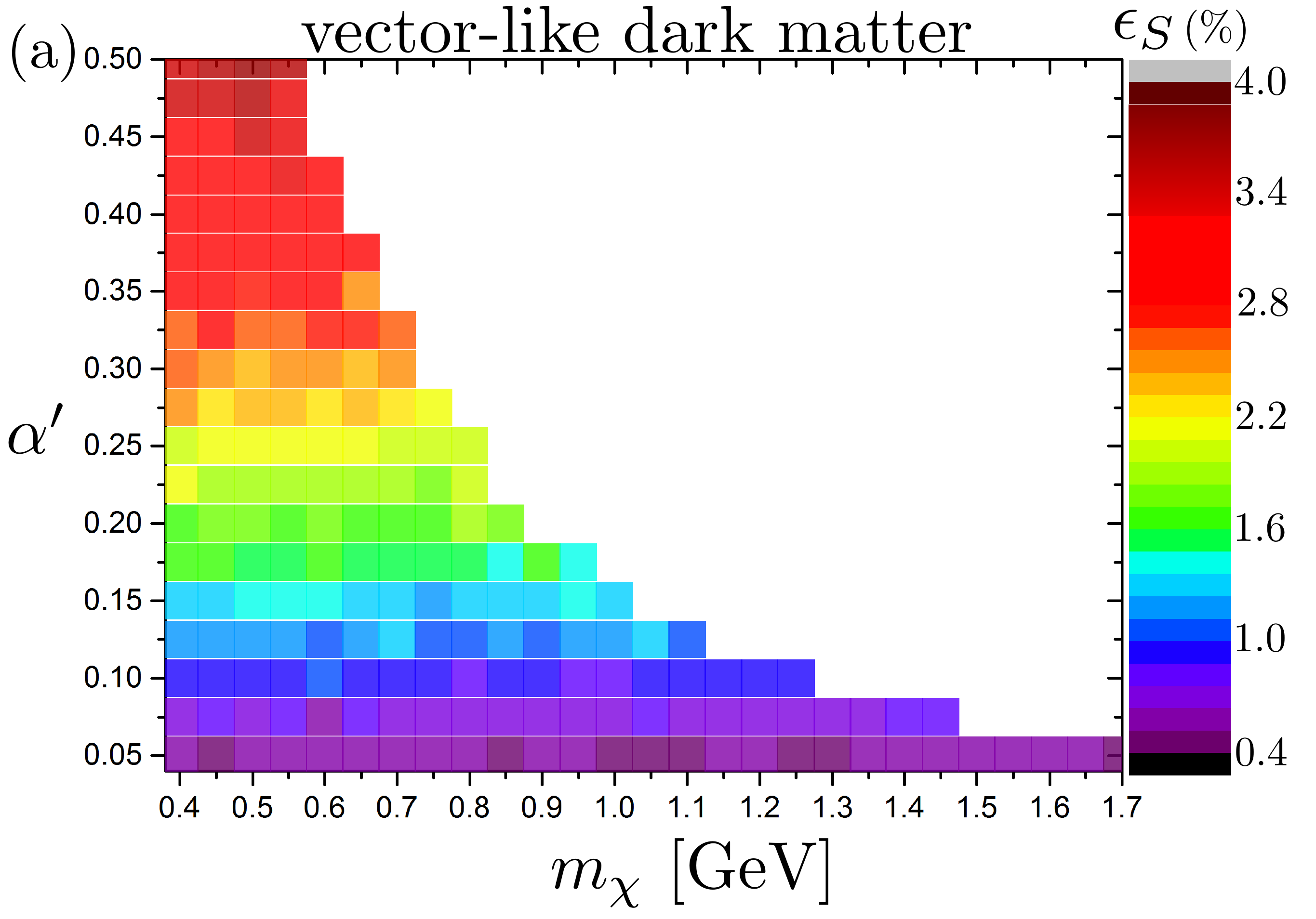} \,\qquad
\includegraphics[width=0.4\textwidth]{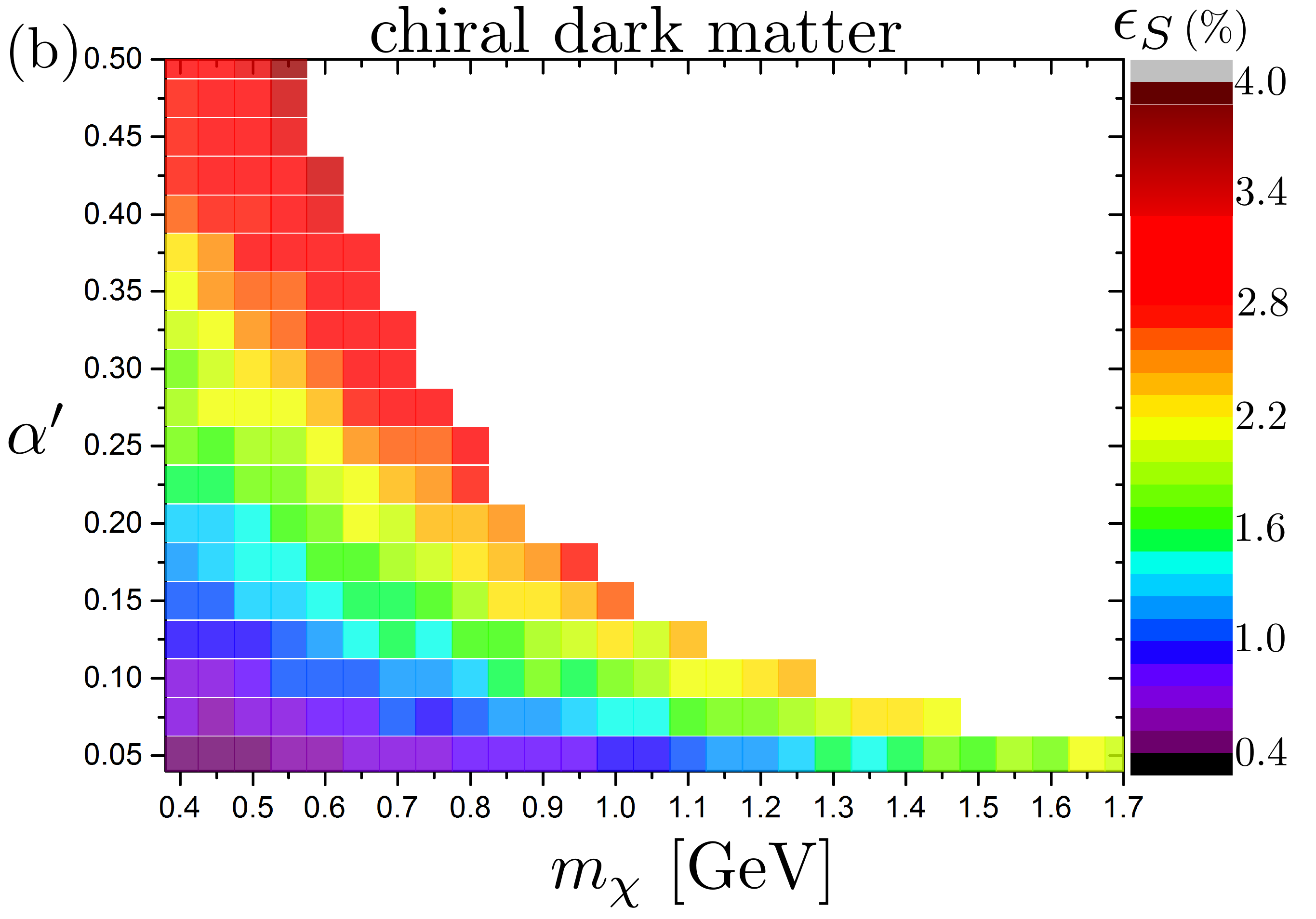} 
\caption{We show the signal efficiency $\epsilon_S$ of tagging two muon-jets as in a search of prompt decaying light bosons of ATLAS\,\cite{Aad:2015sms}.
}
\label{fig:eff}
\end{figure*}
In this section, we present constraints on our benchmark scenarios provide by the LHC searches. 
As we introduce a heavy leptophobic $Z'$ as the mediator of a simplified model for a collider phenomenology,  there is a limit from a heavy resonance search in a dijet signature\,\cite{Aaboud:2017yvp,CMS:2017xrr}. 
The difference between two LHC searches (ATLAS and CMS) for a heavy $Z'$ is that they have different coupling structure between a mediator and each sector (Standard Model sector and dark sector).
ATLAS search assumes an axial-vector coupling between $Z'$ and quarks, dark matter, motivated by negative results from dark matter direct searches\,\cite{Chala:2015ama},  while CMS analysis takes a vector coupling of a model of $U(1)_B$ associated with gauged baryon number\,\cite{Dobrescu:2013coa}. Different coupling structure can affect cut-efficiency as an angular distribution depends on it. But as the difference is proportional to a ratio, $m_d^2 / M_{Z'}^2$ where $m_d$ is the mass of particles from $Z'$ decaying, we can safely combine results from ATLAS and CMS together in recasting these analyses to our case with $M_{Z'} \gg m_q, m_\chi$. 

To apply constraints from dijet resonance searches, we consider the invisible branch ratio of $Z'$. With $M_{Z'}=1.5\,\textrm{TeV}$, the most stringent upper limit on a coupling $g_q$ between $Z'$ and the Standard Model quarks $q$ is from the ATLAS as $g_q \lesssim 0.07$.
This upper limit comes from narrow-width case where $Z'$ decays only to light quarks. A constraint becomes milder with increasing $Z'$ width 
due to loosing sensitivity in wider dijet mass window\,\cite{Aaboud:2017yvp}. Thus our estimation with a narrow-width $Z'$ is conservative. 
Sizable invisible decay partial width from a coupling $g_\chi$ between dark matter and $Z'$, the limit on $(g_q,\,g_\chi)$ can be imposed by
\begin{equation}
g_q^2 \times \frac{N_c \cdot N_{\textrm{lf}}\cdot g_q^2}{N_c \cdot N_f \cdot g_q^2+g_\chi^2} \lesssim 0.07^2 \times  \frac{N_c\cdot N_{\textrm{lf}}\cdot g_q^2}{N_c \cdot N_f \cdot g_q^2}\, ,
\label{eq:dijet}
\end{equation}
with $N_c=3$ is the color factor of the Standard Model and $N_{\textrm{lf}}=5$ is the number of light flavor quarks considered as the final state jets in the LHC dijet searches. 
Here $N_f=5+\sqrt{1-4 m_t^2/M_{Z'}^2} \simeq 5.97$ is the effective number of quark flavors contributing to the width of $Z'$\,\cite{CMS:2017xrr}. 
As we see in the left side of eq.\,\eqref{eq:dijet}, the constraint from dijet searches become weaker as $Z'$ has non-negligible invisible decay proportional to $g_\chi^2$. As the (partial) decay width of $Z'$ is not sensitive to the mass of sub-GeV dark matter, a constraint from dijet searches will not depend on $m_\chi$. 
\begin{table}[h!]
\begin{tabular}{|c|c|c|c|c|}\hline 
Benchmark case  & $\alpha'$ & $m_{\chi} (\textrm{GeV})$ & Model & $\epsilon_S (\%)$  \\
\hline 
(a) strong coupling limit &  0.50 & 0.5 & vector-like & 3.68 \\
\hline 
(b) weak coupling limit & 0.05 & 0.45 & chiral & 0.41\\
\hline
\end{tabular} 
\caption{Benchmark points to consider ATLAS prompt lepton-jet (LJ) analysis conservatively. For (a) strong coupling limit, we take a vector-like dark matter with $(m_\chi,\alpha') = (0.5\GeV, 0.5)$ and for (b) weak coupling limit, we choose a chiral dark matter with $(m_\chi,\alpha') = (0.45\GeV, 0.05)$. Here $\epsilon_S$ is a signal efficiency in tagging at least two isolated muon-jets ($\textrm{LJ}_\mu$) from the ATLAS\,\cite{Aad:2015sms}.}
\label{Accep}
\end{table}

We also consider constraints from prompt lepton-jet analysis\,\cite{Aad:2015sms}. Out of various combinations of lepton-jets as a signal channel, we consider two muon-jets signal as this channel has the highest tagging efficiency and lowest backgrounds\,\cite{Aad:2015sms,Aad:2012qua}. In recasting a prompt lepton-jet analysis, we take two extreme cases where (a) the parameter space allows the maximum dark photon showering activities, and (b) we have least dark photon showering activities within the perturbative limit of $\alpha' \,m_\chi^2 / m_{\gamma_d}^2 \lesssim 1$ as in Tab.\,\ref{Accep}. These two limiting cases provide the upper and lower bounds of the signal efficiency $\epsilon_S$ in tagging two isolated muon-jets as in FIG.\,\ref{fig:eff}. Thus we can see the maximally and minimally allowed range of $g_\chi$ from
the prompt lepton-jet analyses in our parameter space of $(\alpha',\, m_\chi)$ by considering these two extreme cases.

\subsection{Analysis at the High Luminosity (HL) LHC}

In this section, we provide a viable parameter space to have enough statistics in discriminating two different hypothesis on the property of dark matter. 
For the HL-LHC analysis, we consider a muon-jet as following. First of all,  a candidate muon in the muon-jet needs to have $p_T > 5\GeV$ within $|\eta| < 2.5$.
Then we use the Cambridge/Aachen jet algorithm\,\cite{CA} with FastJet\,\cite{Cacciari:2011ma} to cluster a muon-jet with $\Delta R < 0.1$. An isolation variable for a muon-jet is defined as:
\begin{equation}
\rho = \frac{\sum_i E_{T,i}}{p_{T,\mu J}}\, ,
\end{equation}
with $i$ running over all the $E_T$ deposit in the calorimeter, without the candidate muons, near the muon-jet with $\Delta R < 0.3$, and the denominator is the $p_T$ of the muon-jet.
To suppress the QCD faking rate, we require the isolation criteria as $\rho < 0.3$. After that, the  major backgrounds in our case where we require at least two prompt muon-jets would be low-mass mesons including $\rho, \omega$ and $\phi$ which decay into muon pair\,\cite{Aad:2015sms,Aad:2012qua}
\footnote{In the parameter space of non-promptly decaying light particles, $b\bar b$ production becomes a major background as $b$-quarks decay into pairs of muons via double semileptonic decays\,\cite{CMS:2016tgd}.}.
To reduce these backgrounds, we require the mass of reconstructed muon-jet within $(0.3,\,0.5)$ in the unit of [GeV]. This requirement reduces backgrounds from low-mass mesons, leaving the di-photon process ($pp\to \gamma^* , \gamma^* \to  2\,\textrm{LJ}_\mu$) as the major background.
\begin{figure}[t!]
\centering
\includegraphics[width=0.4\textwidth]{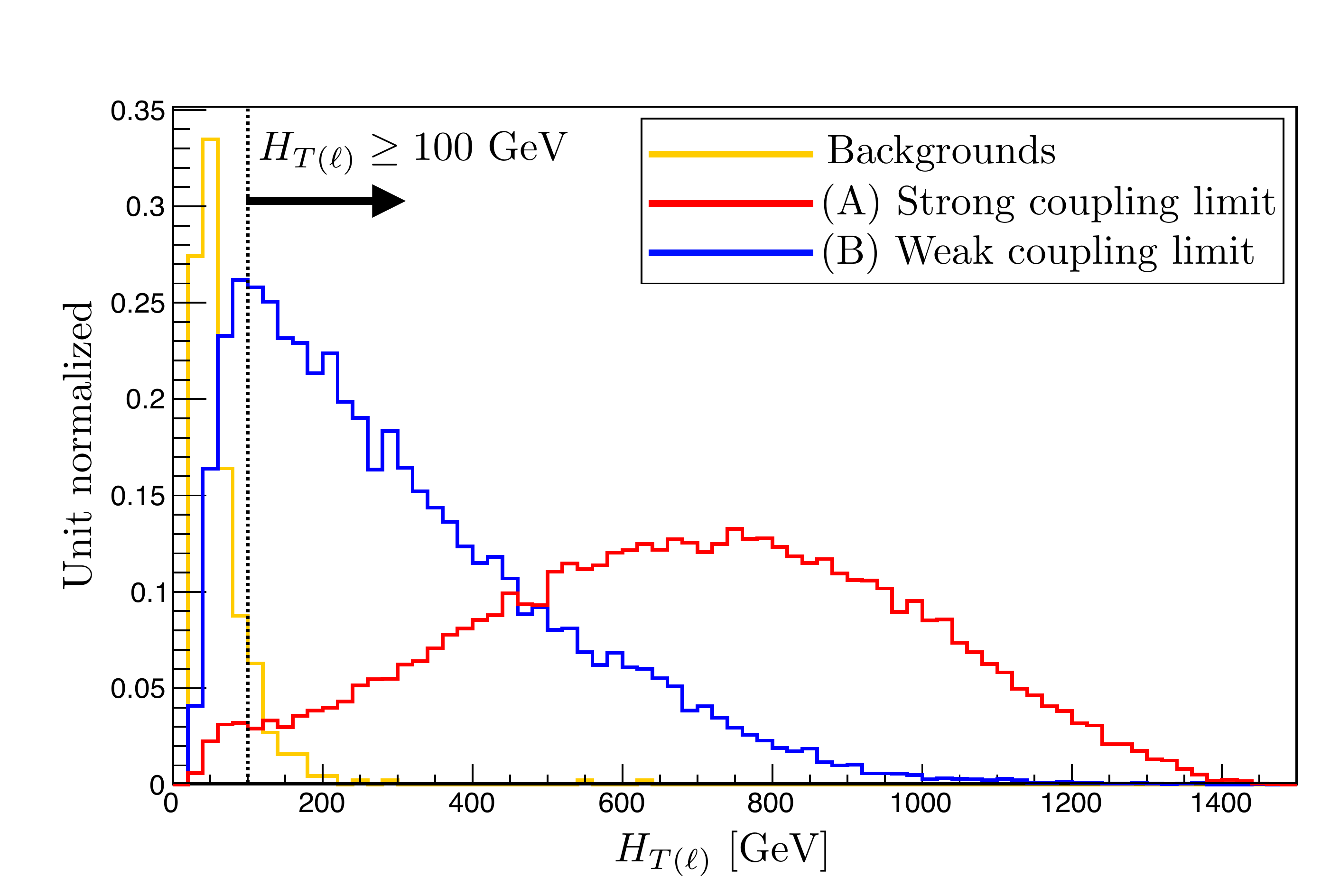} 
\caption{Distributions of $H_{T(\ell)}$ after requiring isolation and mass window on reconstructed moun-jets. The dominant background is from the di-photon production where off-shell photons $\gamma^*$ decays into muon pair. 
The background distribution is shown with a yellow line. we require a cut on $H_{T(\ell)}$ to suppress backgrounds accordingly for the HL-LHC study.
}
\label{fig:HT}
\end{figure}
\begin{figure*}[th!]
\centering
\includegraphics[width=0.4\textwidth]{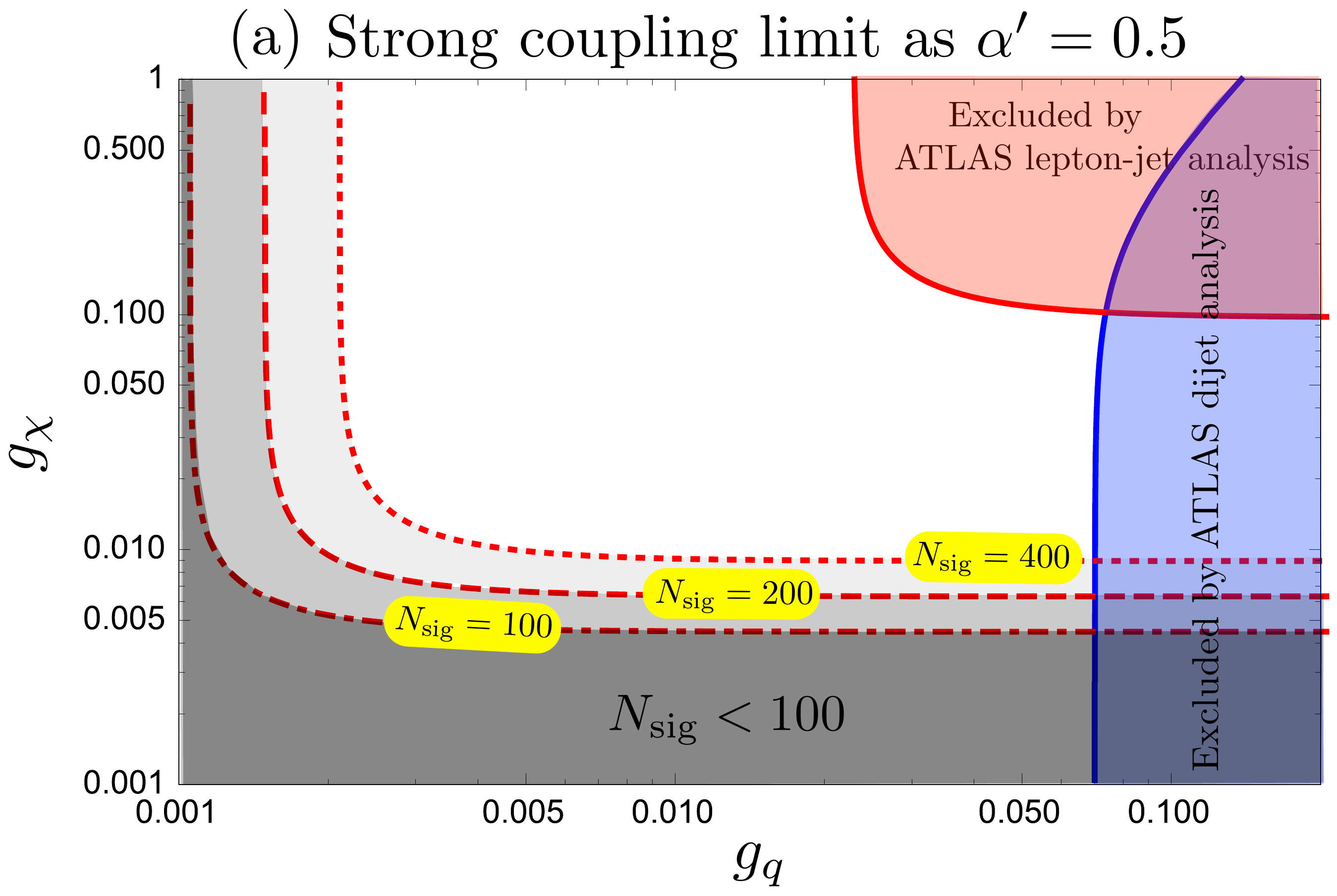} \,\qquad
\includegraphics[width=0.4\textwidth]{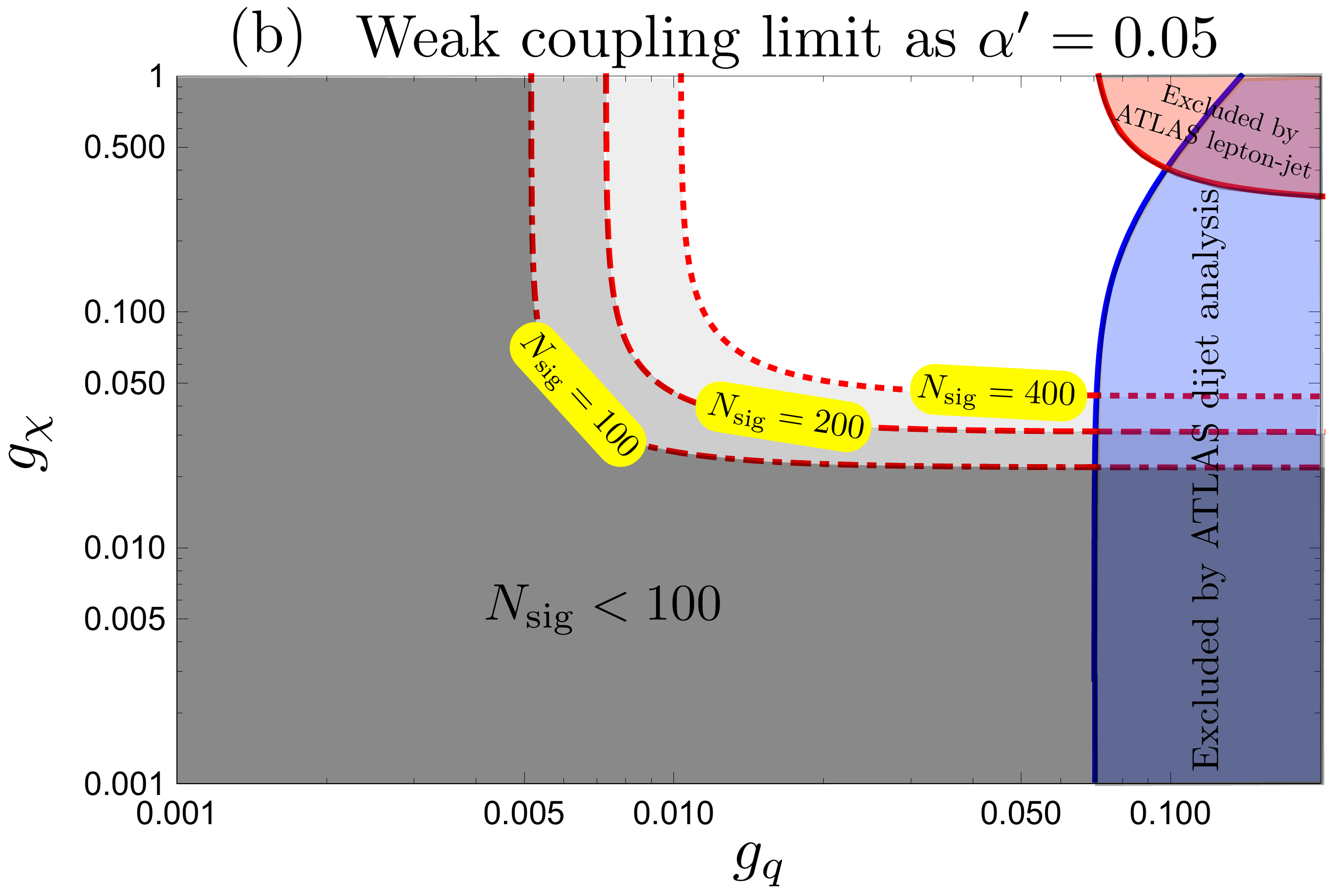} 
\caption{We show possible parameter space of $(g_q, \,g_\chi)$ with moderate events after analysis cuts at the HL-LHC of $\mathcal{L}=3\,\textrm{ab}^{-1}\,$to discriminate chiral dark matter from vector-like dark matter. The blue shaded region is excluded by the ATLAS dijet searches. The red region is excluded by the ATLAS prompt lepton-jet search. 
The grey shaded regions are divided by the number of signal events $N_\textrm{sig}$ less than 100, 200 and 400 after analysis cuts.
}
\label{fig:HLLHC}
\end{figure*}
To suppress contribution from a di-photon process, we require $H_{T(\ell)} \ge 100\GeV$. As we see in FIG.\,\ref{fig:HT}, the most of events in background locate at low mass region of $H_{T(\ell)}$. By requiring a cut on $H_{T(\ell)}$, we have only 3.07 events from backgrounds at 14\,TeV HL-LHC.
\begin{figure*}[th!]
\centering
\includegraphics[width=0.3\textwidth]{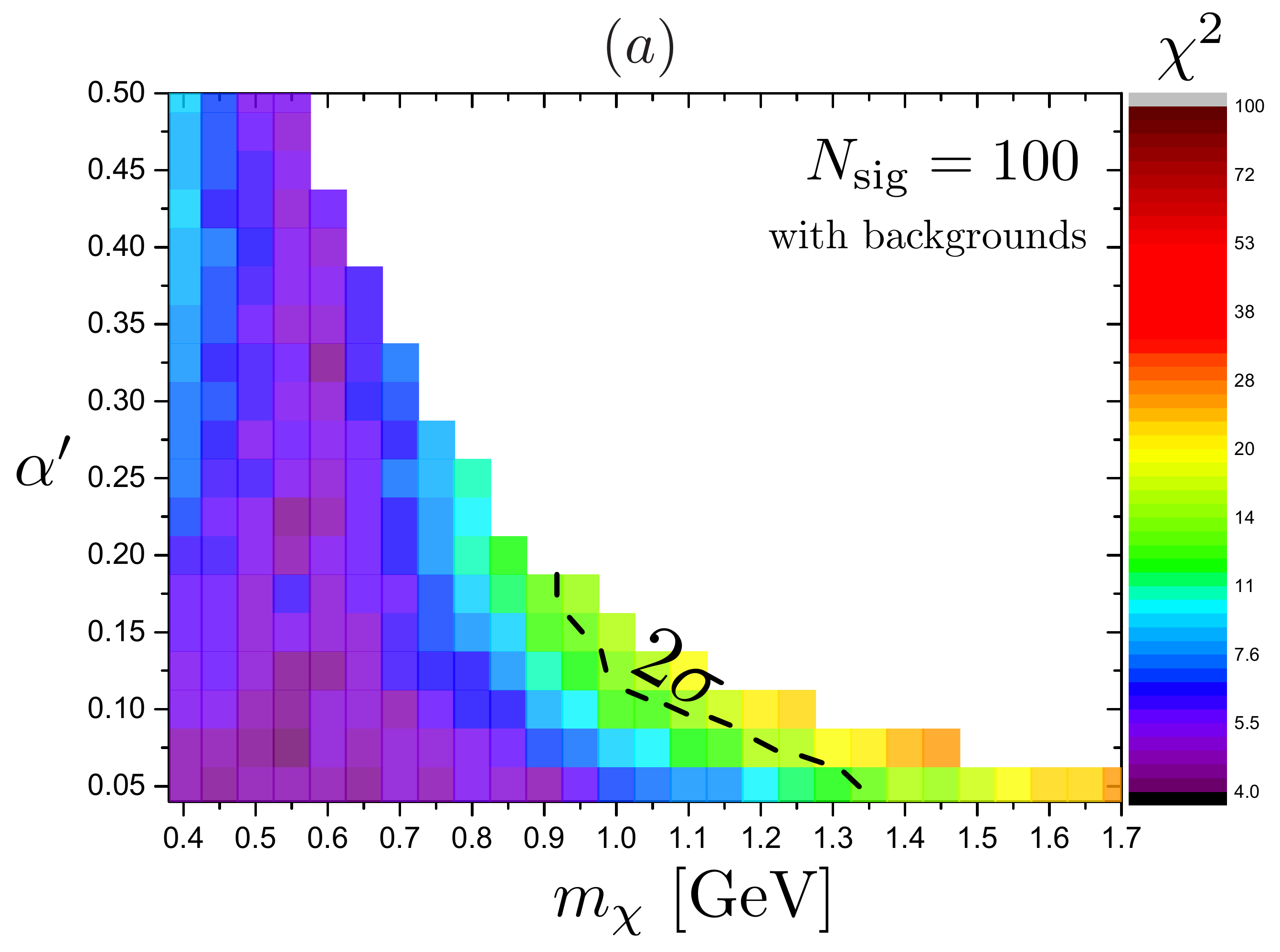} \,\qquad
\includegraphics[width=0.3\textwidth]{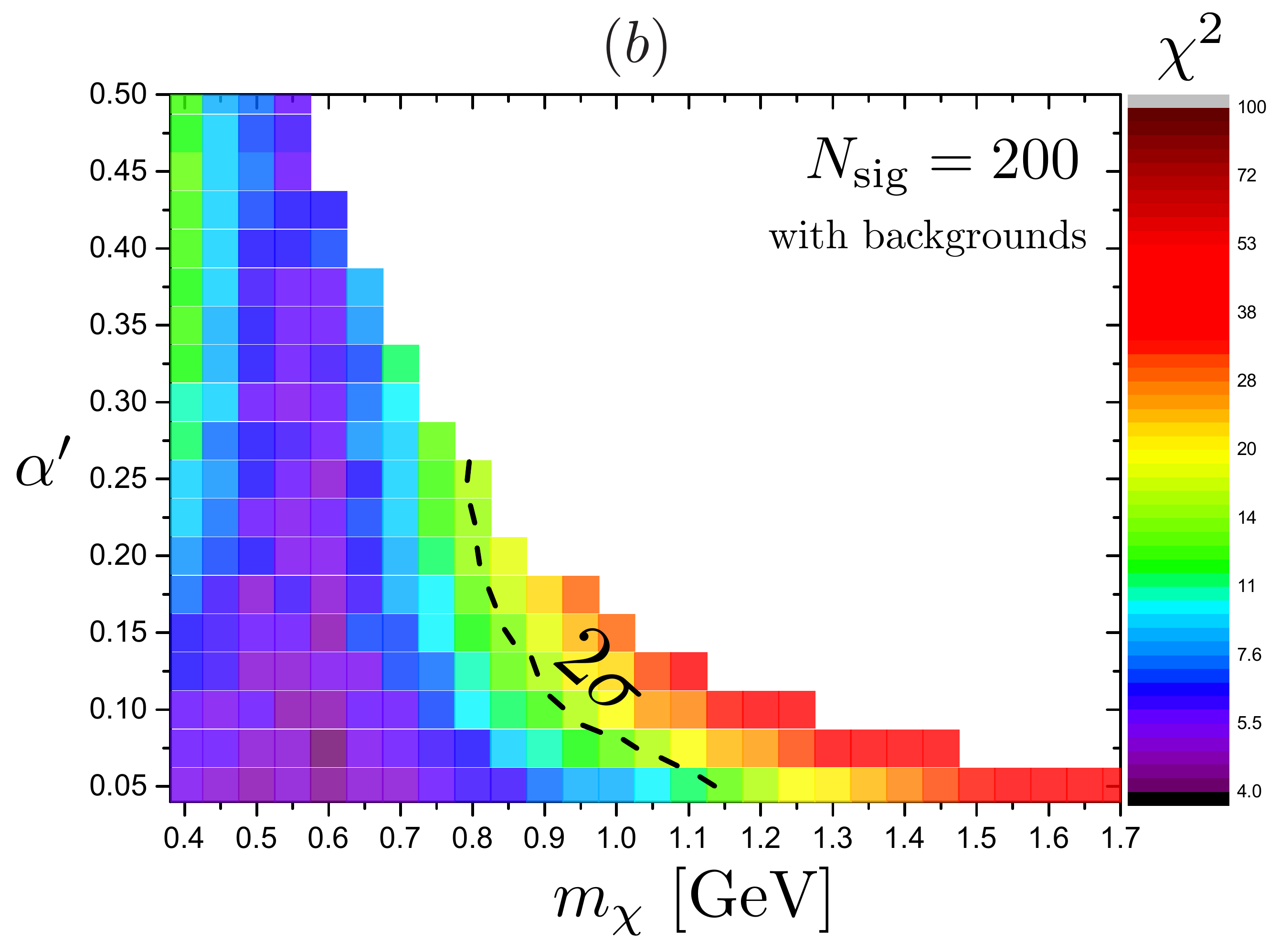} \,\qquad
\includegraphics[width=0.3\textwidth]{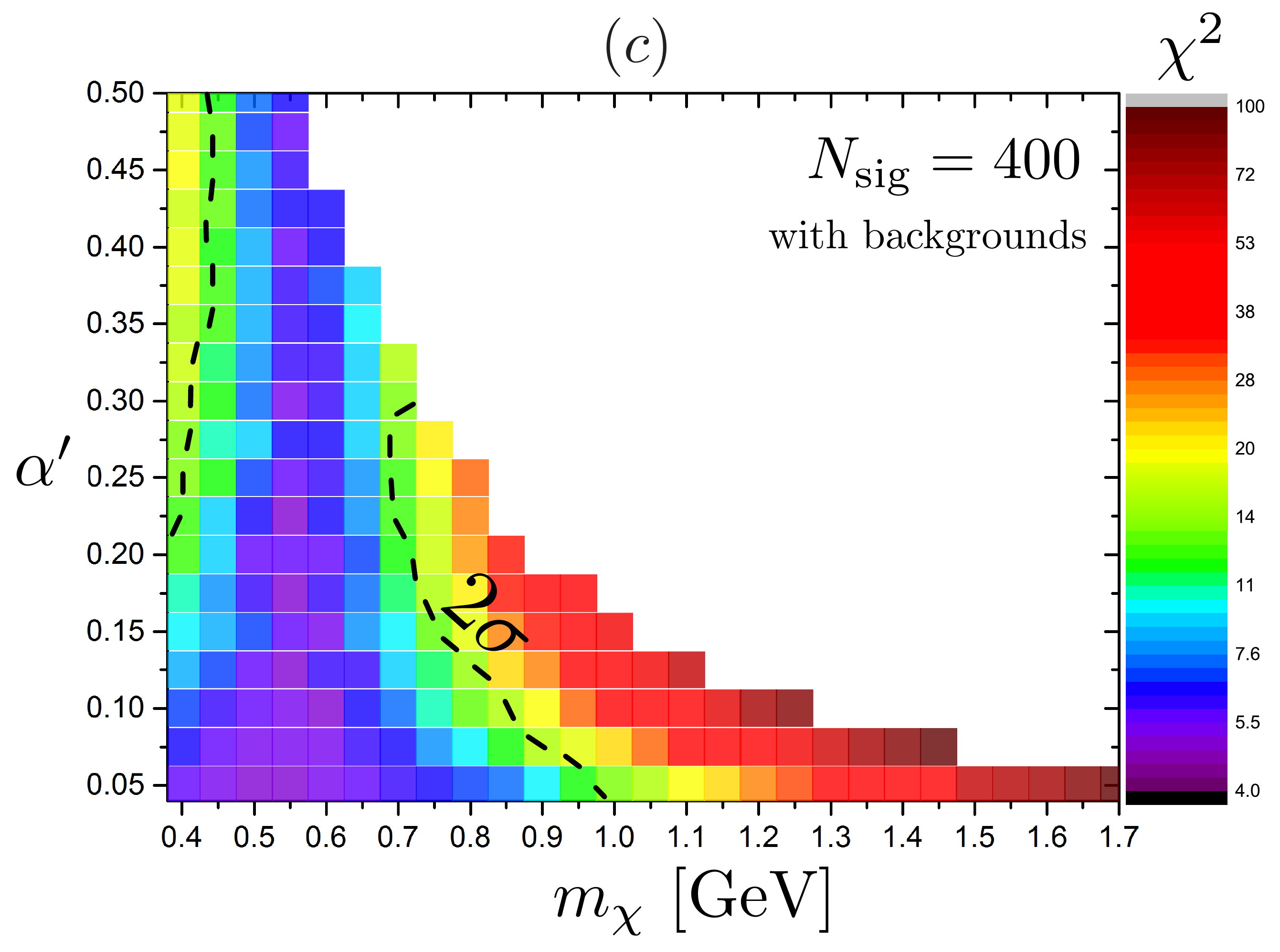} 
\caption{$\chi^2$ distance of $H_{T(\ell)}$ distributions between a vector-like dark matter scenario and chiral dark matter case with 100, 200, and 400 signal events with backgrounds after cuts.}
\label{fig:chi2}
\end{figure*}
We show the parameter space of $(g_q,\,g_\chi)$ which has not been excluded by current LHC analyses and will have enough statistics to distinguish vector-like and chiral dark matter at the HL-LHC of $\mathcal{L}=3\,\textrm{ab}^{-1}$ in FIG.\,\ref{fig:HLLHC}.
Constraints from dijet analysis does not depend on $\alpha'$ as the analysis focuses on dijet final states. It becomes weak as $g_\chi$ becomes large enough to induce a significant invisible decay partial width of $Z'$. 
The number of events $N_\textrm{sig}$ depends on $\alpha'$ as the signal tagging efficiency depends on the amount of dark photon showering which is proportional to $\alpha'$ in vector-like dark matter case. As it is shown in FIG.\,\ref{fig:HLLHC}, there would be viable parameter space where we can see the relation between the mass of dark matter and the mass of dark photon by examining the dark photon shower pattern. 
Finally,  we describe how one can understand the nature of a dark sector once we observe dark matter signatures at the LHC. 
For a collider observable to identify the nature of dark matter, we compare ${H_T}_{(\ell)}$ variable in eq.\,(\ref{HTsum}) distribution with muons of $p_T\ge 5\GeV$ and electrons of $p_T\ge 10\GeV$ by considering momentums from tracks and energy deposits in sub-detectors. To suppress backgrounds which populate the low $H_{T(\ell)}$ region, we consider  only high mass region of $H_{T(\ell)} \ge 100\GeV$.
In our simulation, we have $\mathcal{O}(1)$ events from backgrounds at HL-LHC. 
To deal with a finite luminosity of the LHC, we generated 200 reconstructed signal events after cuts. 
We perform 100 pseudo-experiments to reduce statistical uncertainties coming from finite statistics of simulations.
We calculate a binned-$\chi^2$ of $H_{T(\ell)}$ distributions from a vector-like case and chiral case with including corruptions from backgrounds as we described above.
Fig.\,\ref{fig:chi2} shows our results from $\chi^2$ comparison. The LHC can tell the origin of the mass of a dark matter particle by discriminating a vector-like and chiral dark matter models more than $2\sigma$ significance level, 
when mass ratio $m_{\chi}/m_{\gamma_d}$ is large enough. 
While with a significant events number, as in Fig.\,\ref{fig:chi2} (c), for parameter region of large $\alpha'$ and small $m_{\chi}/m_{\gamma_d}$, 
the vector-like and chiral dark matter models can be also distinguished by $2\sigma$ significance level because of the weaker transverse modes shower in the chiral dark matter case.

\section{Discussions and outlook.}
The Standard Model had been developed by constructing gauge structures. At the LHC, we checked its validity from the discovery of the Higgs particle which provides a mass to SM particles.
Similarly identifying the gauge structure of a dark sector together with understanding the mass origin of dark particles would be the first step toward expanding the physics of dark matter once we observe a dark matter signature. 
With the performance of the LHC which is a complementary tool to probe a light dark matter, we study the feature of collider signatures 
from a dark photon showering depending on the property of dark matter under a dark gauge group.  With numerical simulations of $14\TeV$ LHC, we show that we can identify the nature of dark matter with $\mathcal{O}(100)$ signal events at high energy colliders.
\begin{acknowledgments}
MZ appreciates helps from authors of Pythia 8, Torbjorn Sjostrand, Nishita Desai and Peter Skands for modifying a Pythia showering subroutine.
This work is supported by IBS (Project Code IBS-R018-D1), NRF Strategic Research Program (NRF-2017R1E1A1A01072736), and NRF Basic Science Research Program (NRF-2017R1C1B5075677)
\end{acknowledgments}


\end{document}